# Long-lived Valley Polarization of Intra-Valley Trions in Monolayer WSe$_2$

Akshay Singh[1], Kha Tran[1], Mirco Kolarczik[2], Joe Seifert[1], Yiping Wang[1], Kai Hao[1], Dennis Pleskot[3], Nathaniel M. Gabor[3], Sophia Helmrich[2], Nina Owschimikow[2†], Ulrike Woggon[2] and Xiaoqin Li[1*]

[1] Department of Physics and Center for Complex Quantum Systems, University of Texas at Austin, Austin, TX 78712, USA.
[2] Institut für Optik und Atomare Physik, Technische Universität Berlin, Berlin 10623, Germany
[3] Department of Physics and Astronomy, University of California, Riverside, CA 92521, USA
E-mail address: *elaineli@physics.utexas.edu
† nina.owschimikow@physik.tu-berlin.de

We investigate valley dynamics associated with trions in monolayer tungsten diselenide (WSe$_2$) using polarization resolved two-color pump-probe spectroscopy. When tuning the pump and probe energy across the trion resonance, distinct trion valley polarization dynamics are observed as a function of energy and attributed to the intra-valley and inter-valley trions in monolayer WSe$_2$. We observe no decay of a near-unity valley polarization associated with the intra-valley trions during ~ 25 ps, while the valley polarization of the inter-valley trions exhibits a fast decay of ~ 4 ps. Furthermore, we show that resonant excitation is a prerequisite for observing the long-lived valley polarization associated with the intra-valley trion. The exceptionally robust valley polarization associated with resonantly created intra-valley trions discovered here may be explored for future valleytronic applications such as valley Hall effects.

The valley degree of freedom (DoF) indexes the crystal momentum of a local energy minimum within the electronic band structure, and has been proposed as an alternative information carrier, analogous to charge and spin [1]. In atomically thin transition metal dichalcogenides (TMDs), fundamental optical excitations, excitons (electron-hole pairs) and trions (charged excitons), are formed at the hexagonal Brillouin zone boundaries at the $K$ ($K'$) points. As such, they inherit the valley index which is locked with electron spins in TMDs. Thus, exciton and trion resonances allow optical access and manipulation of the valley DoF in TMDs using circularly polarized light [2-6]. The exceptionally large binding energies of these quasiparticles (i.e. 200-500 meV for excitons and an additional binding energy of 20-40 meV for trions) further promise room temperature valleytronic applications [2,3,7-13].

High efficiency valley initialization and a long lifetime of valley polarization are preferred in valleytronic applications [14-17]. Initial experiments based on steady-state photoluminescence have shown the possibility of creating a near-unity valley polarization in MoS$_2$ and WSe$_2$ via exciton resonances [4,18]. Time-resolved measurements soon revealed that exciton valley polarization is quickly lost (~ 1 ps) due to intrinsic electron-hole exchange interaction [19]. The large initial exciton valley polarization observed in the steady-state PL results from the competition between the valley depolarization time (~ 1 ps) and the exciton population relaxation time (~ 100-200 fs) [13,20,21]. On the other hand, trions offer an interesting alternative route for optical manipulation of the valley index for a number of reasons. First, in contrast to the ultrafast exciton population relaxation time, trions exhibit an extended population relaxation time of tens of picoseconds in monolayer TMDs [22-30]. Secondly, trions as charged quasiparticles influence both transport and optical properties of TMDs and may be readily detected and manipulated in experiments such as valley Hall effect [31]. Last but not the least, previous studies of negatively charged trions in conventional doped semiconductors suggest that negatively charged trions leave the background electron gas spin-polarized after the electron-hole recombination [32-36]. Thus, trions may play a particularly important role in manipulating electron spins and the valley DoF.

In this report, we investigate valley polarization dynamics associated with negatively charged trions in monolayer WSe$_2$ using polarization resolved two-color pump-probe spectroscopy with sub-nm spectral resolution. Distinct valley polarization dynamics were observed as the resonant pump/probe energy is tuned across the trion resonance and attributed to the two types of trions known to exist in monolayer WSe$_2$, intra-valley and inter-valley trions. In particular, we discover a long-lived near-unity valley-polarization (>> 25 ps) associated with the resonantly created intra-valley trions. This exceptionally robust valley polarization (in comparison to excitons and inter-valley trions) originates from the peculiar requirement of *simultaneous* transfer of three carriers (two electrons and one hole) to the other valley with proper spin and crystal momentum changes. When the pump energy is tuned to the exciton resonance, the long-lived trion valley polarization dynamics can no longer be observed, highlighting the difficulty in assessing intrinsic trion valley dynamics under non-resonant excitation conditions used in the majority of previous experiments [3,30,37]. The discovery of an exceptionally robust trion valley polarization is significant since it suggests that information encoded in the valley index can be stored and manipulated electrically via effects such as valley Hall effect over long time scales.

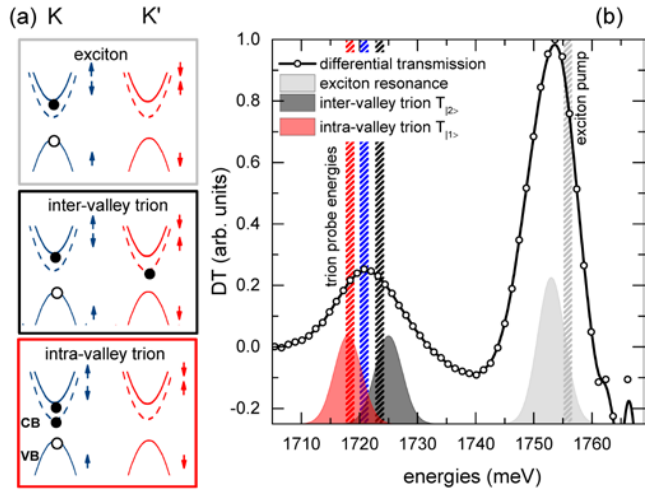

Figure 1: (a) Possible configurations of charged and neutral bright excitons in WSe$_2$. (b) Differential transmission spectrum of WSe$_2$ (dots). The black line is an interpolation curve, serving as a guide to the eye. The solid Gaussians illustrate the spectral position of the exciton and the two trion (inter and intra-valley) resonances. The spectral positions of probe energies for data in Figure 2 and 3 (dashed colored lines), and the pump energy for Figure 3 (gray line) are also illustrated.

In monolayer WSe$_2$, the particular band structure and optical selection rules suggest that the energetically lowest inter-band transition is dipole-forbidden [38-40] as illustrated in Fig. 1a. Thus, two types of negatively charged trions, intra-valley and inter-valley trions may form, represented with symbols $T_{|1\rangle}$ and $T_{|2\rangle}$, respectively. The two electrons have the opposite (same) spins for intra-valley trions $T_{|1\rangle}$ (inter-valley trions $T_{|2\rangle}$). Because of the different spin configurations, the diagonal exchange interaction present in the inter-valley trions $T_{|2\rangle}$ lifts the energy degeneracy and leads to a shift of ~ 6 meV relative to the intra-valley trions $T_{|1\rangle}$ as illustrated in Fig. 1a [41,42]. The exciton resonance is approximately 30 meV higher than $T_{|2\rangle}$, which has implications for optical phonon assisted scattering between $T_{|2\rangle}$ and exciton resonances [43,44].

We study a mechanically exfoliated monolayer WSe$_2$ flake on a sapphire substrate (kept at temperature ~ 13 K) using a pump/probe set up briefly described here (See the supplementary for an image of the sample, a photoluminescence spectrum, and a schematic of the nonlinear experiment [19]). The sample is considered to be *n*-doped based on similarly prepared samples from previous studies [18,42]. The output from a mode-locked Ti-sapphire laser is split into pump and probe beams, whose wavelengths are independently varied by two grating-based pulse shapers. After the pulse shapers, the pulse duration is ~ 1 ps (with ~ 0.7 nm bandwidth). After passing through linear polarizers and $\frac{1}{4}\lambda$ waveplates for circular polarization control, the beams are focused to a spot size of ~ 2 $\mu$m. The power for each beam is kept at ~ 10 $\mu$W to obtain nonlinear signals in the $\chi^{(3)}$ regime and to avoid heating effects. The transmitted differential transmission (DT) signal is detected following further spectral filtering through a spectrometer, which allows us to study trion dynamics under resonant excitation conditions. DT is defined as, $DT = (T_{pump\ on} - T_{pump\ off})/T_{pump\ off}$, where $T_{pump\ off(on)}$ is the transmitted probe intensity when pump is off (on), and it measures the third order nonlinear response.

We first performed a fully degenerate experiment using cross-linearly polarized pump/probe beams to identify exciton and trion resonances at 1753 meV and 1719.6 meV respectively as shown in Fig. 1b. We note that intra-valley and inter-valley trions are not spectrally resolved in our sample as those on BN substrates [44]. Nevertheless, both types of trions are intrinsic to WSe2 and should be present under the inhomogeneously broadened trion resonance.

We then investigate the trion valley dynamics by simultaneously tuning the pump/probe energy across the trion resonance. The pump energy is kept at 2 meV above the probe energy to allow filtering of the scattered pump after passing through the spectrometer. This quasi-resonant excitation condition is referred to as the resonant excitation condition in this paper for simplicity. In the following, a $\sigma^+$ polarized pump pulse is used to populate the $K$ valley and the subsequent dynamics in the $K$ ($K'$) valley is investigated using a $\sigma^+$ ($\sigma^-$) polarized probe pulse. The co- and cross-circularly polarized DT signals are displayed in the same panel as a function of time delay between the two pulses as shown in Fig. 2a-c. The co-circular experiments reflect trion population relaxations within the same valley and have similar features in all scans: after an initial rise during the excitation pulse (solid grey area), there is a relatively fast decay of a few picoseconds followed by a slower decay of ~ 35 ps. The observed bi-exponential decay is consistent with previous experiments and likely arises from scattering between the bright trion states and dark states (or trap states) [19,22].

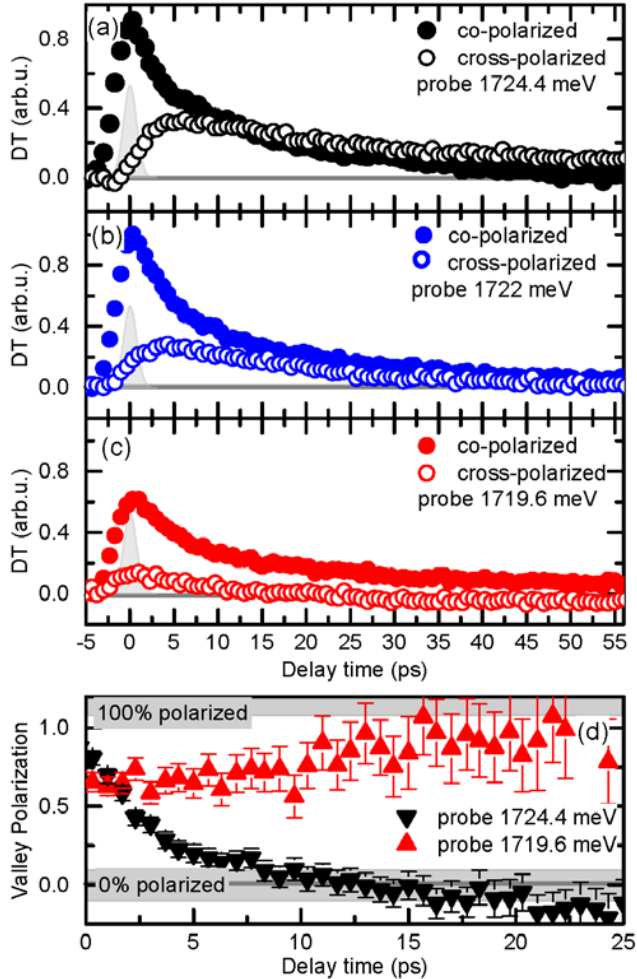

Figure 2: Co- (solid dots) and cross-polarized (open dots) signals recorded in resonant excitation experiments across the trion resonance, at probe energies of (a) 1726.8 meV, (b) 1722 meV, and (c) 1719.6 meV; (d) Valley polarization for the measurements displayed in (a) and (c).

The most intriguing feature is the drastic and systematic change in the cross-circularly polarized scans as the pump/probe energies are tuned through the trion resonance as shown in Fig. 2a-c. In cross-circularly polarized experiments, trions created in the $K$ valley are converted to trions in the $K'$ valley via spin flip and electron-hole exchange interaction. We attribute the dynamics on the higher (lower) energy side of the trion resonance to inter-valley trions $T_{|2\rangle}$ (intra-valley trions $T_{|1\rangle}$). For inter-valley trions $T_{|2\rangle}$ probed at 1724.4 meV, the population in the opposite valley builds up and reaches its maximum value after a few picoseconds (Fig. 2a) [45]. In contrast, valley scattering is minimal for intra-valley trions $T_{|1\rangle}$ probed at 1719.6 meV during trion population relaxation time as shown in Fig 2c. The robust valley polarization associated with $T_{|1\rangle}$ is reflected in the minimal cross-circularly polarized signal shown in Fig. 2c. As the excitation energy is tuned further to the lower energy, negative DT signal appeared only for the cross-circularly polarized scans (data shown in Ref [19]). This negative DT signal for cross-circularly polarized scan likely arises from valley-dependent many body effects [25,46,47]. We limit the following discussion to the spectral region with only positive DT signal where the valley polarization can be defined meaningfully.

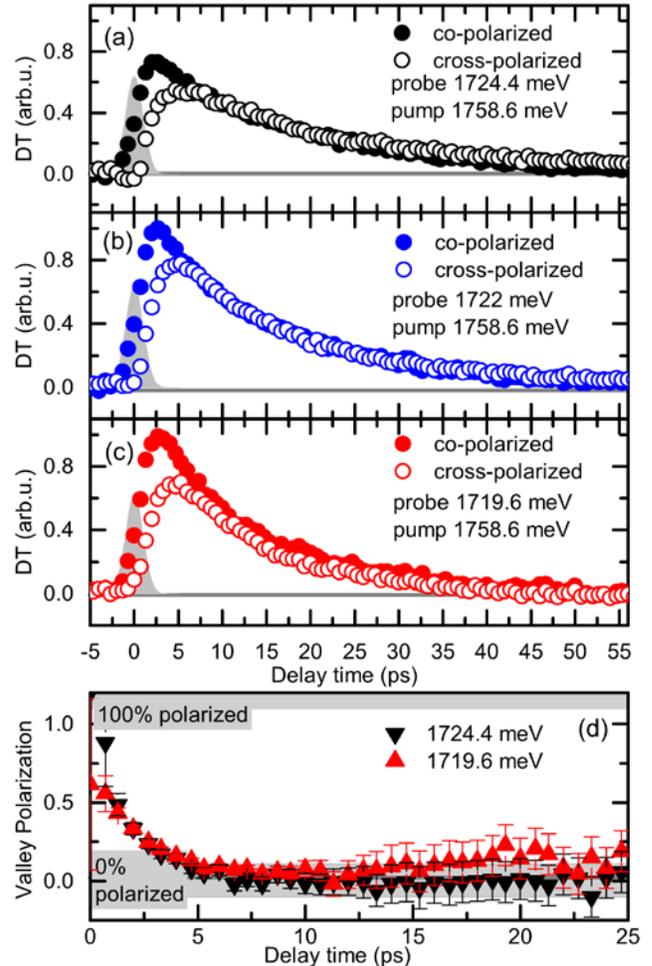

Figure 3. Co- (solid dots) and cross-polarized (open dots) signals recorded in non-resonant excitation experiments for pumping at the exciton resonance and probing at (a) 1724.4 meV, (b) 1722 meV, and (c) 1719.6 meV; (d) Valley polarization for the measurements displayed in (a) and (c).

We define valley polarization as $VP = (co - cross)/(co + cross)$, following earlier work on TMDs [4,48] and calculate trion valley polarization for two particular probe energies at 1724.4 meV and 1719.6 meV, respectively. We focus on these two energies to highlight the distinct trion valley dynamics associated with the two types of trions, while minimizing spectral overlap between them. Trion valley polarization at these two energies as a function of time delay between the pump/probe is shown in Fig. 2d. The valley polarization is only plotted over a limited delay range because the error bars become very large at larger delays due to the small DT signal in both the co- and cross-circularly polarized scans [49]. The inter-valley trion valley polarization $T_{|2\rangle}$ exhibits a fast decay of ~ 4 ps, which are consistent with earlier studies [22]. In contrast, the valley polarization associated with the intra-valley trion $T_{|1\rangle}$ persists much longer, and decays with a time constant much larger (> 25 ps) than the experimental observation range. A valley depolarization time longer than the population relaxation time associated with the intra-valley trions means that these trions recombine before valley scattering occurs, leaving the residual electron valley/spin polarized.

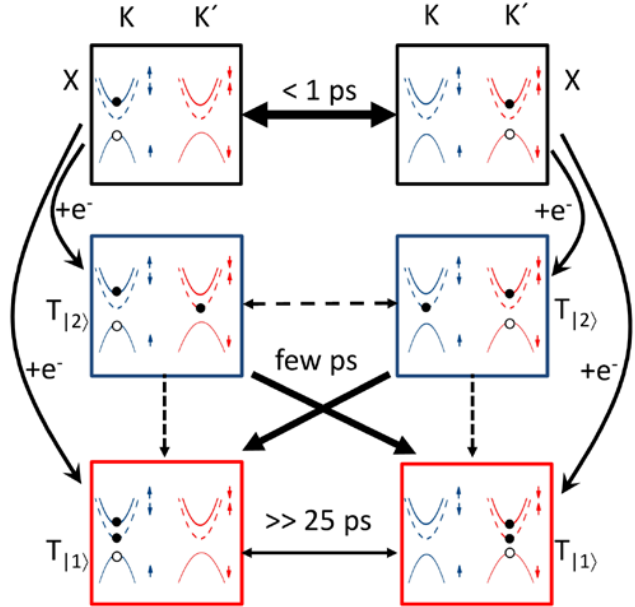

Figure 4: Schematic of the suggested valley polarization dynamics, exciton and trion conversion processes, and their respective time scales as measured in the experiment. Dashed lines suggest that such processes are possible in principle, but do not compete favorably with other faster processes.

Furthermore, this long-lived trion valley polarization associated with $T_{|1\rangle}$ is only observable under resonant excitation conditions. When we excited the mobile excitons at the higher energy side of the exciton resonance (1758.8 meV specifically) while tuning the probe energy within the trion resonance, the difference between valley polarization dynamics for $T_{|1\rangle}$ and $T_{|2\rangle}$ disappears as shown in Fig. 3a-c. An apparent fast valley depolarization (~ 2 ps) is observed for probe energy tuned to both types of trions as shown in Fig. 3d. These experiments performed under the non-resonant excitation conditions do not report on the intrinsic trion valley dynamics. Instead, it is necessary to consider a number of physical processes including the valley depolarization of excitons, trion formation, and phase space filling in the interpretation. The key feature of similar and rapid valley depolarization for probing at both trions mainly arises from the rapid exciton valley depolarization, i.e., excitons created at the $K$ valley quickly scatters to the $K'$ valley within the pulse duration (~ 1 ps) due to electron-hole exchange interaction [19,36,50]. The same DT signal amplitude in the co- and cross-circularly polarized experiments after 5 ps support the interpretation of equal trion populations at the two valleys [51]. In the co-circular experiments, the DT reaches its maximal value immediately after the excitation pulse. The creation of excitons at the $K$ valley prohibits the formation of either type of trions in the same valley due to phase space filling, leading to an instant and reduced absorption at the trion energy. In the cross-circular experiments, the finite DT signal rise time (1-2 ps) is determined by the time for the exciton to capture an extra charge, i.e., the trion formation time [52]. These experiments unequivocally illustrate the importance of near-resonant excitation to access the intrinsic dynamics associated with the trion valley DoF.

We summarize the various exciton and trion conversion and valley dynamics in a diagram shown in Fig. 4. The top of the diagram illustrates the rapid exciton valley depolarization (~ 1 ps and shorter than the excitation pulses used in our experiments) due to electron-hole exchange interaction. Trion valley depolarization is expected to be slower than that associated with excitons because it requires an additional carrier spin flip. Interestingly, the drastically different valley polarization dynamics associated with the two types of trions in WSe2 have never been explicitly proposed or observed experimentally. The $K$ valley $T_{|2\rangle}$ can scatter to the opposite valley and form $K'$ valley $T_{|2\rangle}$ without loss of energy. This process, however, is not as efficient as scattering to $K'$ valley $T_{|1\rangle}$. The latter process occurs through electron-hole exchange interaction and is energetically favorable. Thus, we suggest that this $K$ valley $T_{|2\rangle}$ to $K'$ valley $T_{|1\rangle}$ conversion process is responsible for the ~ 4 ps inter-valley trion valley depolarization observed. Inter-valley trions created in the $K$ valley can also be converted to intra-valley trion (the vertical dashed arrow) in the same valley via a spin flip, which is likely a slower process as illustrated by the vertical dashed lines. We discuss inter- to intra-valley trion conversion process in more detail in Ref [19]. Finally, intra-valley trion valley depolarization is long-lived as illustrated in the bottom of the diagram. The transfer of either a single electron or an electron-hole pair to the other valley transforms the intra-valley trion into an inter-valley trion, which is an energetically unfavorable process. Scattering of $K$ valley $T_{|1\rangle}$ to the opposite valley requires the *simultaneous* transfer of three carriers (two electrons and a hole) to the other valley. Thus, valley polarization of the intra-valley trions in monolayer WSe2 is exceptionally stable, consistent with our experimental observations. Valley polarized PL from the trion resonance was previously observed under non-resonant excitation conditions in MoS2 [3]. In addition to being different TMD materials, various time scales (population relaxation, valley depolarization, and trion formation) are manifested differently in PL and DT experiments. Systematic studies are necessary to investigate how these time scales vary among different TMD samples placed on various substrates at different doping levels.

Microscopic theory of valley dynamics associated with trions with different spin configurations and exchange interaction are not available yet. The experiments presented here provide further motivation and challenges for such theoretical studies on valley dependent exchange interaction and many body effects due to Coulomb interaction, which is particularly pronounced in monolayer semiconductors. Most importantly, this work suggests a possible approach for creating and manipulating long-lived valley DoF potentially useful in valleytronic applications.

**Acknowledgements**: We gratefully acknowledge helpful discussions with Galan Moody and Andreas Knorr. The spectroscopic experiments were jointly supported by NSF DMR-1306878 (A. Singh) and NSF EFMA-1542747 (K. Tran, K. Hao, J. Seifert, and X. Li). The collaboration on sample preparation between UT-Austin (K. Tran) and UC-Riverside (D. Pleskot and N.M. Gabor) was supported as part of the SHINES, an Energy Frontier Research Center funded by the U.S. Department of Energy (DOE), Office of Science, Basic Energy Science (BES) under Award # DE-SC0012670. Tran, Li, Pleskot, and Gabor have received support from the SHINES. Li also gratefully acknowledges the support from a Humboldt fellowship, which facilitated the collaboration with TU-Berlin. Kolarczik, Helmrich and Woggon acknowledge funding from Deutsche Forschungsgemeinschaft via GRK 1558.